# Should we adjust for pupil background in school value-added models?

# A study of Progress 8 and school accountability in England

George Leckie and Harvey Goldstein

Centre for Multilevel Modelling and School of Education, University of Bristol

**Address for correspondence**

Centre for Multilevel Modelling

School of Education

University of Bristol

35 Berkeley Square

Bristol

BS8 1JA

United Kingdom

g.leckie@bristol.ac.uk

# Should we adjust for pupil background in school value-added models?
# A study of Progress 8 and school accountability in England

**Summary.** In the UK, US and elsewhere, school accountability systems increasingly compare schools using value-added measures of school performance derived from pupil scores in high-stakes standardised tests. Rather than naïvely comparing school average scores, which largely reflect school intake differences in prior attainment, these measures attempt to compare the average progress or improvement pupils make during a year or phase of schooling. Schools, however, also differ in terms of their pupil demographic and socioeconomic characteristics and these also predict why some schools subsequently score higher than others. Many therefore argue that value-added measures unadjusted for pupil background are biased in favour of schools with more 'educationally advantaged' intakes. But, others worry that adjusting for pupil background entrenches socioeconomic inequities and excuses low performing schools. In this article we explore these theoretical arguments and their practical importance in the context of the 'Progress 8' secondary school accountability system in England which has chosen to ignore pupil background. We reveal how the reported low or high performance of many schools changes dramatically once adjustments are made for pupil background and these changes also affect the reported differential performances of region and of different school types. We conclude that accountability systems which choose to ignore pupil background are likely to reward and punish the wrong schools and this will likely have detrimental effects on pupil learning. These findings, especially when coupled with more general concerns surrounding high-stakes testing and school value-added models, raise serious doubts about their use in school accountability systems.

# 1. Introduction

In the UK, US and elsewhere, education systems increasingly hold schools to account using school performance measures derived from pupil scores in high-stakes standardised tests and examinations (NFER, 2018; OECD, 2008; Koretz, 2017). Schools are held accountable for the progress or improvement shown by their pupils over a year or phase of schooling. The implicit assumption is that variation in school average progress is a valid indicator of the value that schools add to pupil learning. In other words, the education effectiveness or quality of schools.

England has been at the forefront of this move to test-based school accountability (West, 2010). Successive governments over the last twenty-five years have introduced new and supposedly improved school performance measures that purport to measure what is happening in schools (Kelly and Downey, 2010; Leckie and Goldstein, 2017). These measures are also used to promote parental choice via their high-profile publication in 'school league tables' (Leckie and Goldstein, 2009). They are also used by schools for self-evaluation, improvement, tracking, and target setting purposes, with schools increasingly buying in data analysis support from commercial organisations to assist them in these endeavours (Selfridge, 2018, p.40). The measures also inform national debates around regional inequalities, the performance of different school types, and performance gaps across socioeconomic, ethnic, and other pupil groups.

In 2016, the Government introduced a new secondary school accountability system for all mainstream stated-funded schools in England (DfE, 2018c). Attainment 8 – essentially a total score across eight traditional academic subjects – was introduced as the new headline measure of pupil performance at the end of secondary schooling General Certificate of Secondary Examinations (GCSEs; age 15/16). Progress 8 – a type of value-added approach – was introduced as the new headline measure of progress or the improvement that pupils make between the end of primary schooling key stage 2 tests (KS2; age 10/11) and the GCSE examinations. Each pupil's score is calculated as their Attainment 8 score minus the average Attainment 8 score of all pupils nationally with the same KS2 prior attainment (KS2 scores are categorised into 34 groups for this purpose). A school's Progress 8 score is simply the average of their pupils' scores and is presented with a 95% confidence interval to communicate its statistical uncertainty. It can be shown that the statistical modelling approach underlying Progress 8 is therefore a two-stage linear regression approach as opposed to the usual multilevel (random-effects) modelling approach used in the school effectiveness research literature, a point we return to later. The Government argue Progress 8 leads to fairer and more meaningful comparisons for school accountability purposes than Attainment 8 as it adjusts for school intake differences in KS2 prior attainment. Specifically, schools are labelled 'underperforming' if their Progress 8 scores fall below a minimum standard for progress referred to as a 'floor standard'. Such schools come under increased scrutiny and intervention from Ofsted, the national school inspectorate, and by regional schools commissioners and local authorities in their roles supporting schools. In contrast, schools with the highest Progress 8 scores are exempt from routine inspections by Ofsted in the following calendar year, a highly desirable outcome for any school.

The design of all school value-added measures and accountability systems is based on subjective modelling decisions and assumptions and given the high stakes nearly always involved, these choices must be independently and robustly evaluated. In this article, we explore a particularly divisive decision relevant to not just Progress 8, but all measures and systems, namely whether to adjust for school intake differences in pupil demographic and socioeconomic background characteristics since these factors also predict why some schools subsequently score higher than others. We assess the practical importance of this decision for Progress 8 and school accountability where the Government has chosen to ignore pupil background. We examine in detail the extent to which schools' Progress 8 scores, ranks and classifications as successful and failing schools change when we account for pupil background. We highlight those schools which would benefit and lose most by any change to Progress 8. We then draw attention to further statistical issues with Progress 8 which demand further research as well as our reservations more generally with regard to test-based school accountability.

**2. To adjust or not to adjust?**

Progress 8 adjusts pupils' Attainment 8 scores for their KS2 prior attainment scores but does not adjust for other pupil characteristics which also differ across schools. While prior attainment is nearly always the most important predictor of current attainment in school value-added models, many national and international studies have long shown the secondary importance of pupil demographic and socioeconomic characteristics as additional predictors (Reynolds et al., 2014; Teddlie and Reynolds, 2000). It follows that, in absence of any adjustments, different pupil groups will typically show different average progress during schooling. Thus, in England, girls typically make more progress during secondary schooling than boys, many ethnic minority groups make more progress than White British pupils, pupils with no special education needs make more progress than those with needs, and rich pupils make more progress than poor pupils (EPI, 2017). It follows that schools with more 'educationally advantaged' intakes in England would in general be expected to show higher average pupil Progress 8 scores than schools with less educationally advantaged intakes. Some studies have additionally shown that school average pupil prior attainment and various school average background characteristics also predict subsequent pupil attainment even after adjusting for the pupil versions of these variables (Timmermans and Thomas, 2015), but we do not consider this further in this article.

*The argument for adjusting: To make fair and meaningful comparisons*

Many academics and educationalists argue that failing to adjust for pupil background is fundamentally unfair as it punishes some schools merely for teaching educationally disadvantaged intakes and rewards other schools merely for teaching educationally advantaged intakes and (BBC, 2018; Goldstein, 1997; OECD, 2008; Raudenbush and Willms, 1995; Reynolds et al., 2014; Teddlie and Reynolds, 2000; TES, 2018). The true effectiveness of many schools in disadvantaged areas will go undetected as will the lack of effectiveness of many schools in advantaged areas. School value-added measures such as Progress 8 which ignore pupil background are therefore likely to punish and reward the wrong schools and to hold up the wrong schools as examples of success that other schools should learn from. Furthermore, punishing schools for teaching disadvantaged pupils is likely to incentivise schools to avoid admitting particular pupil groups (e.g., children with special educational needs), and where they are admitted, to find ways to exclude them from the examinations and therefore the value-added calculations. Indeed, in England, there has been a large rise in pupil exclusions over the last two years which in part has been attributed to schools gaming the accountability system in these ways (DfE, 2018a). A related concern is that unadjusted school value-added measures require disadvantaged pupils in each school to make as much progress as their advantaged peers. However, given the differential performance of many pupil groups, this is simply an unrealistic target, at least in the short run, and so is likely to leave many disadvantaged pupils and their schools feeling as if they have failed. This may dissuade good teachers from working in challenging schools and may induce teachers in those schools to leave. Proponents of all these arguments therefore argue that school value-added measures must adjust not just for prior attainment but additionally for pupil socioeconomic status and other pupil characteristics that predict subsequent attainment.

*The argument against adjusting: It lowers expectations of disadvantaged groups*

Others argue against adjusting school value-added measures for pupil background, worrying that such adjustments entrench socioeconomic inequities and excuse low performing schools. In terms of Progress 8, the UK Government argues that society should expect disadvantaged pupils with the same prior attainment as their more advantaged peers to continue to perform at the same academic level at GCSE, not fall behind (Burgess and Thomson, 2013; DfE, 2010). There is, however, a lack of any theoretical justification for such an assertion. Moreover, it seems inconsistent to acknowledge the empirical fact that pupils from disadvantaged backgrounds are already behind when they start their secondary schooling, but to refuse to accept the empirical fact that this 'deficit' is not fully removed by adjusting for their lower prior attainment.

The Government go on to argue that adjusting for the lower progress of disadvantaged pupil groups entrenches low aspirations for these pupils (DfE, 2010). However, if one accepts this argument then one must also accept that adjusting for prior attainment entrenches low aspirations for low prior attaining pupils. Thus, using this argument to ignore pupil background but to adjust for pupil prior attainment appears inconsistent (Perry, 2016).

One practice that clearly entrenches low aspirations for particular pupil groups is the widespread practice of target setting in schools since here empirical relationships between attainment and pupil background characteristics in previous school cohorts is used to predict the future performance of current pupils and so past inequities are passed onto future generations (Ho and Castellano, 2013; Leckie and Goldstein, 2017; Selfridge, 2018). However, the importance placed on target setting in England and elsewhere is driven by the high-stakes nature of school accountability systems, and thus is questionable on those grounds, rather than implying an underlying flaw in adjusting for pupil background.

**3. Data**

We focus on the 3,098 schools whose Progress 8 scores were published in the Government's 2016 secondary school performance tables: essentially all state maintained secondary schools in England. We use the Government's National Pupil Database to recreate the underlying pupil-level Attainment 8 and KS2 score dataset from which school Progress 8 scores are derived. We additionally merge in a range of standard pupil background and school characteristics. Pupil characteristics include: age, gender, ethnicity, language (whether they speak English as an additional language), SEN (special educational needs status), FSM (eligible for free school meals at some time in the preceding six years: an indicator of poverty), and deprivation (deprivation of the pupil's residential neighbourhood as proxied by the IDACI decile of their home postcode). School characteristics include: region, type, admissions policy, age range, gender, religious denomination, and deprivation (deprivation of the school neighbourhood). The final analysis sample consists of 502,851 pupils in 3,098 schools located in 151 local authorities across the nine regions of England.

Table 1 presents pupil- and school-level summary statistics for Progress 8. See the Supporting Information (Figures S1-S3) for pupil- and school-level summary statistics and plots for Attainment 8, KS2 prior attainment, and Progress 8. A 1-unit difference in Progress 8 corresponds to a 1 grade difference per GCSE subject. Pupil Progress 8 scores are approximately normally distributed with a national mean and SD of 0 and 1.06. The mean and SD of school average Progress 8 scores are -0.03 and 0.40 and its distribution is also approximately normal.

Table 2 presents school Progress 8 'bandings'. Essentially, the Government assigns each school to one of five bands as a function of the magnitude and statistical significance of their Progress 8 score (DfE, 2018b; see Table 2 for the exact definition of each banding). We see that 303 schools nationally (9.8% of all schools) are assigned to the 'well below average' banding and therefore do not meet the Government's minimum standard of progress (defined as the threshold between this banding and the 'below average' banding). In contrast, 193 schools nationally (6.2%) are assigned to the 'well above average' banding.

**4. The relationship between Progress 8 and pupil background characteristics**

In this section, we reveal the very different average pupil progress made by different pupil groups according to Progress 8. Figure 1, left-hand panel, presents average pupil Progress 8 by pupil age, gender, ethnicity, language, SEN, FSM, and deprivation. The categories within each pupil characteristic are sorted by average pupil Progress 8 and for each pupil characteristic the overall variation across the categories is statistically significant (one-way ANOVA tests robust to school-level clustering all show $p < 0.001$). These statistics are preliminary descriptive statistics which analyse each pupil characteristic separately. Later, we will model pupil progress jointly in terms of all seven characteristics. See Supporting Information for the number of pupils across the categories of each pupil characteristic (Table S2) and for corresponding plots for Attainment 8 and KS2 prior attainment (Figure S4).

August born pupils make 0.19 grades more progress per subject than their September born peers. Given that the SD in pupil Progress 8 is 1.06, this difference is substantial, almost one fifth of 1 SD. More generally, younger pupils within the academic year make more progress than older pupils. However, younger pupils score lower than older pupils at the end of primary schooling and they still do so at the end of secondary schooling despite their higher progress (Supporting Information: Figure S4). Thus, the higher progress shown among younger pupils reflects their attainment approaching, but not reaching, the higher attainment of their older peers during secondary schooling. These patterns agree with Crawford et al. (2013) and others who have done work on month of birth effects in England.

Girls make more 0.26 grades more progress per subject than boys. However, girls already score higher than boys at the end of primary schooling (Supporting Information: Figure S4) and so the end of primary school gender attainment gap widens over secondary schooling. Potential explanations are discussed in detail by Sammons (1995) among others.

There is substantial variation in Progress 8 by ethnic group. Chinese pupils (0.3% of all pupils) score, on average, 0.70 grades higher per subject than expected given their prior attainment, Indian pupils (2.5%) 0.49 grades higher, Black African pupils (2.9%) 0.37 grades higher, and Bangladeshi pupils (1.5%) 0.35 grades higher. In contrast, White British pupils (76%), on average, score 0.08 grades lower than expected. Black Caribbean pupils (1.3%) do worse still, scoring 0.11 grades lower than expected. However, Gypsy/Roma pupils (0.1%) and Travellers of Irish Heritage (0.02%) show the lowest progress, scoring 0.64 and 1.04 grades lower. These progress gaps in England are long-standing and their causes are complex and intertwined with the differing socioeconomic status and other characteristics of these groups (Strand, 2014; Wilson et al., 2011).

Pupils speaking English as an additional language (13% of all pupils) make 0.48 grades more progress per subject than pupils who speak English as their first language. Essentially, this pupil group catches up and by the end of secondary schooling overtakes their peers who speak English as a first language (Supporting Information: Figure S4). Strand et al. (2015) describe in detail the relationships between pupil attainment, progress and language status in England.

Pupils with SEN support (11% of all pupils), especially those with statements (2%), make considerably less progress than pupils with no special education needs. These two pupil groups already score lower at the end of primary schooling and so these attainment gaps widen during secondary schooling (Supporting Information: Figure S4).

Pupils eligible for FSM (27% of all pupils) make 0.43 grades less progress per subject than pupils who are not eligible for FSM. Ilie et al. (2017) provide a recent discussion of FSM differences in progress including the strengths and weaknesses of using FSM as a proxy for socioeconomic disadvantage.

Pupils residing in disadvantaged neighbourhoods also make less progress than those in more prosperous neighbourhoods. For example, pupils living in the most affluent 10% of neighbourhoods score, on average, 0.19 grades higher per subject than predicted by their prior attainment, while pupils living in the poorest 10% of neighbourhoods score 0.27 grades lower per subject than predicted. This social gradient is already present at the end of primary schooling and so widens over secondary schooling (Supporting Information: Figure S4).

**5. Modifying Progress 8 to adjust for pupil background characteristics**

In this section, we modify Progress 8 to adjust for the seven pupil background characteristics described above: age, gender, ethnicity, language status, SEN, FSM, and deprivation. We refer to this measure as 'Adjusted Progress 8'.

Recall that each pupil's Progress 8 score is calculated as their actual Attainment 8 score minus the average Attainment 8 score across all pupils nationally with the same KS2 prior attainment, where KS2 prior attainment is categorised into 34 bands for this purpose. The calculation of pupil and school Progress 8 scores can therefore be viewed as an application of linear regression. Essentially, pupil Progress 8 scores are calculated as the residuals from a linear regression of pupil Attainment 8 on 34 dummy variables, one for each KS2 band. School Progress 8 scores are then calculated as school averages of these residuals. This reformulation reveals the Government's approach to be at odds with the considerable methodological and applied research literature on measuring school effects which favours a multilevel modelling approach, a point we return to in the Discussion (Aitkin and Longford, 1986; Goldstein, 1997, 2011; OECD, 2008; Raudenbush and Willms, 1995; Reynolds et al., 2014; Teddlie, Reynolds, 2000).

We explore the importance of adjusting for pupil background on Progress 8 as simply as possible by entering these seven pupil characteristics into the Progress 8 linear regression model. Thus, we retain all other features of the Government's methodology. We do not include interaction terms as the use of the 34 dummy variables for prior attainment means that interactions between prior attainment and the pupil characteristics would result in a very large number of parameters, many of which would be poorly estimated. Given the importance of accounting for such interactions (Goldstein, 1997), this is a clear limitation of the Progress 8 methodology (it would seem preferable to enter prior attainment as a low order polynomial). Figure 1 (right-hand panel), confirms that the Adjusted Progress 8 model fully adjusts for the seven pupil characteristics: the average pupil progress for every pupil group is now 0.

The full results for the Progress 8 and Adjusted Progress 8 models can be found in the Supporting Information (Table S4). Here we summarise the overall fit of these two models to the data. The Progress 8 model results in 34 regression coefficients. The adjusted R-squared is 0.570 and so pupils' KS2 scores predict 57% of the variation in their Attainment 8 scores. In contrast, the Adjusted Progress 8 model results in 78 regression coefficients and an increased adjusted R-squared of 0.624. The standard deviation of pupils' progress scores reduces by 6.6% while the correlation between the pupil Adjusted Progress 8 scores and pupil Progress 8 is 0.895. These statistics suggest that while prior attainment is clearly the most important predictor of Attainment 8, the seven pupil characteristics nonetheless improve these predictions.

## 6. Comparing Progress 8 and Adjusted Progress 8 scores, ranks, and bandings

In this section we reveal the practical importance of adjusting for pupil background by comparing Progress 8 and Adjusted Progress 8 scores, ranks and classifications.

Reconsider Table 1. Focussing on the school-level statistics, the means of both variables are effectively zero, but the SD of Adjusted Progress 8 is lower than that for Progress 8 (0.35 vs. 0.40). Thus, school Adjusted Progress 8 scores are in general smaller in absolute value than school Progress 8 scores. The intuition is that Progress 8 overstates the effects schools have on their pupils: part of the measured effects simply reflects school intake differences in pupils' backgrounds.

Figure 2 presents scatterplots of school Attainment 8, Progress 8, and Adjusted Progress 8 scores (first row) and ranks (second row). The Progress 8 against Attainment 8 scatterplots (first column) suggest schools with the best Attainment 8 results tend, but are no means guaranteed, to be the schools where pupils make the most progress (Pearson correlation: $r = 0.75$; Spearman rank correlation: $r_s = 0.77$). The small cluster of schools distinct from the rest (top plot) are grammar schools whose unusual performance we shall return to later. The Adjusted Progress 8 against Attainment 8 scatterplots (second column) show a somewhat weaker relationship ($r = 0.61; r_s = 0.62$) illustrating again that part of what is measured by Progress 8 is school variation in pupil background. The Adjusted Progress 8 against Progress 8 scatterplots (third column) show the strongest associations ($r = 0.91; r_s = 0.89$). However, even here, school performance differs greatly depending on which progress measure schools are judged. This is shown by the substantial number of schools located away from the 45-degree line in the bottom plot. Indeed, changing from Progress 8 to Adjusted Progress 8 would lead 574 schools (19% of all schools in the country) to move up or down the national league table by 500 or more ranks with 110 schools (4%) moving over 1000 ranks. Bearing in mind that there are only around 3000 secondary schools nationally, these changes are very large indeed.

Table 3 presents a cross tabulation of school Progress 8 bandings (rows) and Adjusted Progress 8 bandings (columns). The row percentages present the percentage of schools within each Progress 8 banding that are assigned to each Adjusted Progress 8 banding. The table shows that moving from Progress 8 to Adjusted Progress 8 would lead 988 schools (32% of all schools) to change bandings. Importantly, the number of schools assigned to the 'well below average' banding and therefore judged to be performing below the Government's floor standard would drop from 303 schools (9.8% of all schools) to 196 schools (6.3% of all schools), a decrease of 107 schools, or just over a third. At the other extreme, the number of schools assigned to the 'well above average' banding would decrease from 193 schools (6.2% of all schools) to 148 schools (4.8% of all schools), a decrease of 45 schools, or almost a quarter

The decrease in the number of schools appearing in these two most extreme bandings is consistent with the lower SD reported for school Adjusted Progress 8 scores compared to school Progress 8 scores (0.35 vs. 0.40; Table 1). The intuition is that by setting more realistic expected Attainment 8 scores for pupils, fewer pupils would be deemed to make irregular progress and so fewer schools would be judged to be substantially under- or over-performing and therefore appearing in the two most extreme bandings. However, this is not to imply that no schools would move into the two most extreme bandings under Adjusted Progress 8. Indeed, 16 schools judged 'below average' under Progress 8 would be judged 'well below average' under Adjusted Progress 8 and therefore now inline for Ofsted intervention. The intuition here is that the previously acceptable average pupil progress seen in these schools is no longer acceptable once we learn that these schools disproportionately teach educationally advantaged pupils.

**7. Comparing Progress 8 and Adjusted Progress 8 scores by school characteristics**

In this section, we describe which types of schools would, on average, benefit or lose from any move to adjust Progress 8 for pupil background. We do this by comparing pupil average Progress 8 and Adjusted Progress 8 scores by school region, type, admissions policy, age range, gender, religious denomination, and deprivation.

The left- and right-hand panels of Figure 3 present pupil average Progress 8 and Adjusted Progress 8 scores by each school characteristic in turn. To facilitate comparisons, the categories within each school characteristic, for both measures, are sorted by average pupil Progress 8 scores. In every case, the variation across the categories of each school characteristic is statistically significant (one-way ANOVA tests robust to school-level clustering all show $p < 0.001$). As with Figure 1, these are simple descriptive statistics which analyse each characteristic separately. See Supporting Information for the number of pupils and schools by each school characteristic (Table S3) and for corresponding plots for Attainment 8 and KS2 prior attainment (Figure S5).

According to Progress 8 (left-hand panel), pupils in London schools (431 schools; 14% of all schools) make, on average, the most progress, scoring 0.19 grades higher per GCSE subject than pupils nationally with the same prior attainment. However, under Adjusted Progress 8 (right-hand panel) this 'London effect' halves to just 0.09 grades per subject. Further analysis suggests that while London schools are somewhat disadvantaged by teaching relatively poor intakes (they have relatively high rates of FSM pupils and pupils in deprived neighbourhoods), they are to a much greater extent advantaged by teaching particular ethnic groups who nationally tend to make high progress (in particular, Black Africans, Any Other Ethnic Group, Any Other White Background, Bangladeshi, and Indian). They also teach high proportions of pupils who speak English as an additional language, another high progress pupil group. See Blanden, et al. (2015) and Burgess (2014) for discussions of this 'London effect'. Now consider schools in the North East (152 schools; 5%), the region which shows almost the lowest average pupil progress according to Progress 8, with a score of -0.11. Under Adjusted Progress 8, this score increases to 0.02. Essentially, under Progress 8, schools in the North East are doubly disadvantaged by teaching not just relatively poor intakes, but by also disproportionately teaching white British pupils. Both of these pupil characteristics are associated with below average progress (Figure 1).

There are now a number of different school types in England (Hutchings and Francis, 2017; IPPR, 2017). Average pupil progress for many school types remains approximately the same when we move from Progress 8 to Adjusted Progress 8. However, for some school types, average pupil progress changes markedly. In particular, among converter academies (1320 schools; 43% of all schools), average pupil progress drops from 0.09 to 0.05, while among sponsored academies (560; 18.1%), average pupil progress increases from -0.15 to -0.04. The superior performance of converter academies over sponsored academies is expected as only successful schools (as judged by Ofsted) are allowed to become converter academies while sponsored academies are usually set up to replace under-performing schools. Here the driving factor for the reduction in their apparent difference in performance is that converter academies teach a much lower percentage of poor pupils (20% eligible for FSM) than sponsored academies (40% eligible for FSM). Similarly, the very low average pupil progress seen in both university technical colleges (26 schools; 0.8%) and studio schools (30 schools or 1%) is substantially reduced once the types of pupils who tend to attend these schools is taken into account. Specifically, studio schools are disadvantaged by teaching a high percentage of SEN pupils (33%), while university technical colleges are disadvantaged by teaching a high percentage of boys (76%).

While nearly all schools in England are comprehensive (they do not in theory select on prior attainment), a small number of grammar schools (162 schools; 4.1%) use entrance examinations (House of Commons Education Committee, 2017). Schools in grammar school areas with no entrance examinations are referred to as secondary modern schools (117 schools; 3.5%). In terms of school admissions, according to Progress 8, pupils in Grammar schools score, on average, a considerable 0.33 grades higher per subject than pupils nationally with the same prior attainment. However, under Adjusted Progress 8, the apparent benefit of attending a grammar school is reduced by almost a third: average pupil progress drops from 0.33 to 0.24. Grammar schools are especially advantaged by the low percentage of poor (6.8%) and to a lesser extent SEN pupils (5.6%) they teach, but are also advantaged by disproportionately teaching various high progress ethnic groups. Interestingly, adjusting for pupil background leads secondary modern schools to appear less rather than more effective: average pupil progress drops from -0.05 to -0.09. The intuition for this result is that while secondary modern schools teach a much higher percentage of poor pupils than grammar schools (23.8% vs. 6.8%), they still teach lower percentages of poor pupils than schools nationally (26.6%). Adjusted Progress 8 takes this into account leading to a slight lowering of average pupil progress.

Schools in England also vary somewhat in the age ranges which they teach. Average pupil Progress 8 varies less dramatically by school age range and so we see only relatively small changes in average pupil progress when we move from Progress 8 to Adjusted Progress 8. According to both measures, there is some suggestion that pupils make more progress in schools teaching through to 18 than in schools teaching through to 16. However, more noticeable is the lower progress made by pupils in schools which teach from age 14 onwards. This last group are disproportionately university technical colleges, studio schools and further education colleges, all of whose low progress was noted above.

While nearly all schools in England are mixed-sex, there are a small number of all-girls schools (209 schools; 6% of all schools) and all-boys schools (151 schools; 4%).

Progress 8 suggests pupils in single-sex schools, especially all-girls schools, make more progress than pupils in mixed sex schools. However, average pupil progress in all-girls schools drops from 0.31 to 0.10 when we move from Progress 8 to Adjusted Progress 8. In contrast, the average pupil progress in all-boys schools increases from 0.15 to 0.19 and so the performance of all-boys schools now appears more impressive than that of all-girls schools. The reason for this change is that Adjusted Progress 8 adjusts for pupil gender whereas Progress 8 does not. Nationally, girls outperform boys (Figure 1). Thus, whereas Progress 8 compares girls in all-girls schools to girls and boys nationally, Adjusted Progress 8 only compares girls in all-girls schools to girls nationally. We note that single-sex schools are disproportionately grammar schools whose higher average pupil progress we have already reported.

A minority of schools in England follow a religious denomination (564 schools; 17.6%) (Long and Bolton, 2018). Progress 8 shows pupils in religious schools typically make more progress than those in schools with no religious character. Especially high progress is seen in the small number of Muslim (8 schools), Jewish (11 schools), and Sikh schools (1 school). However, the results for these schools change markedly when we turn to Adjusted Progress 8. In terms of Muslim schools, average pupil progress halves from 0.78 under Progress 8 to 0.36. The intuition for this drop is that these schools teach very high percentages of Indian (49.5%) and Pakistani (37%) pupils who also don't speak English as a first language (80.7%). These characteristics are nationally associated with making high progress (Figure 1). An even more extreme change is shown by the single Sikh school where average pupil progress changes from 0.34 under Progress 8 to -0.19 under Adjusted Progress 8. The large change seen here reflects that this school almost exclusively teaches Indian pupils (86%), one of the very highest progress ethnic groups. The average pupil progress for Jewish schools, on the other hand, changes little. Here an analysis of the underlying data shows that accounting for ethnicity actually raises average pupil progress slightly as Jewish pupils fall under the White British ethnic group which nationally underperforms. However, Jewish schools also teach relatively prosperous intakes and so the net effect is that their average pupil progress is nonetheless lowered when one also additionally accounts for FSM and deprivation.

Finally, the strong relationship between neighbourhood socioeconomic deprivation and pupil progress weakens substantially as we move from Progress 8 to Adjusted Progress 8. This result is not surprising as Adjusted Progress 8 adjusts for the deprivation of each pupil's neighbourhood, and in general most pupils in each school reside in neighbourhoods of similar deprivation to that of their school.

## 8. Discussion

In this article, we have explored whether school accountability systems should adjust for pupil demographic and socioeconomic background characteristics in their school value-added models. We have critiqued the theoretical arguments for and against making these adjustments and examined their practical importance in the context of England's 'Progress 8' secondary school accountability system. Specifically, we modified Progress 8, which only adjusts for pupil prior attainment, to produce an 'Adjusted Progress 8' measure that additionally accounts for seven further pupil characteristics: age, gender, ethnicity, language, SEN, FSM, and deprivation. We then compared Progress 8 and Adjusted Progress 8 in terms of schools' scores, ranks and classifications, and in terms of pupil average scores across a range of school characteristics.

*The impact of adjusting Progress 8 for pupil background*

Our results for Progress 8 show that adjusting for pupil background qualitatively changes many of the interpretations and conclusions one draws as to how schools in England are performing. For example, over a third of schools judged 'underperforming' according to the Progress 8 floor standard would no longer be judged underperforming according to Adjusted Progress 8. More generally, a fifth of schools would see their national league table positions change by over 500 places, which is substantial given there are only around 3000 schools nationally. Pupil FSM and ethnicity prove the most important characteristics to consider. For example, the high average pupil progress seen in London more than halves when we adjust for pupil background and this is principally due to the high proportions of high progress ethnic groups taught in London. In contrast, the low average pupil progress seen in the North East increases substantially after adjustment due to the disproportionately high proportions of poor pupils taught in this region. Other dramatic changes are seen for Grammar schools and faith schools whose high average pupil progress reduces substantially once the educationally advantaged nature of their pupils is taken into account. In contrast, the low average pupil progress seen in sponsored academies increases once the disadvantaged nature of their pupils is recognised.

*Should we adjust Progress 8 for pupil background?*

It seems clear from our results that the higher the proportion of disadvantaged pupils in a school, the more it will effectively be punished for the national underperformance of these pupil groups. It would therefore seem that value-added measures such as Progress 8 which ignore pupil background implicitly view schools rather than Government or society as primarily responsible for these national differences in performance. In contrast, value-added measures such as Adjusted Progress 8 that account for pupil background can be argued to view Government and society rather than schools as primarily responsible for these national differences. The decision to adjust can therefore be seen as a choice between two opposing views. However, there is no need to choose, especially as most would argue that schools, society, and Government bear shared responsibility for the national differences that we see between different pupil groups. In the English context, it would seem that the Government would therefore do better to publish and explain Progress 8 and an adjusted Progress 8 measure side-by-side to present a more informative picture of schools' performances.

*Further methodological concerns with Progress 8*

There are, however, other unusual methodological features to Progress 8 which raise further doubts as to its purported validity. In particular, Progress 8 follows a two-stage linear regression approach. However, the most commonly applied approach in the literature is to use multilevel models (Aitkin and Longford, 1986; Goldstein, 1997, 2011; OECD, 2008; Raudenbush and Willms, 1995; Reynolds et al., 2014; Teddlie, Reynolds, 2000). We would argue that there are notable benefits of the multilevel approach to studying school effects. First, the approach is more robust to the biases which will arise in the presence of any systematic sorting of more advantaged pupils into more effective schools (Castellano et al., 2014). Second, the predicted school effects are so-called 'shrinkage' estimates which pull the estimated value-added scores of small schools towards the national average and therefore discourage unwarranted conclusions being drawn about the effectiveness of those schools where there is insufficient data to be statistical confident in making any such inferences (Goldstein, 2011). Third, the multilevel approach lends itself to the study of 'differential school effects', the notion that schools may make differential progress with different pupil groups (e.g., low prior attainers or particular ethnic groups) (Strand, 2016). Fourth, multilevel value-added models can easily be extended to incorporate separate scores on different academic subjects and across multiple cohorts of pupils (Leckie, 2018) facilitating richer summers of school performance. Fifth, these models can also be adapted to account for the series of schools mobile pupils attend (Leckie, 2009), as opposed to the default approach of naively holding the final school attended accountable for the entirety of these pupils' education.

*More general limitations of using school value-added measures for school accountability*
Importantly, the methodological concerns we have expressed regarding Progress 8 are just a small subset of more general concerns with high-stakes testing and the use of school value-added models in school accountability systems, voiced both by academics (Amrein-Beardsley, 2014; Foley and Goldstein, 2012; Koretz, 2017; Perry, 2016) and society more generally (NAHT, 2018; betterwithoutbaseline.org.uk; morethanascore.org.uk; vamboozled.com). Key concerns are that the tests fail to measure many important aspects of teaching (e.g., pupil engagement, curiosity, an eagerness to learn), lead to a narrowing of the curriculum (e.g., they typically ignore arts, music, drama, and other non-traditional academic subjects), result in teaching to the test, induce excessive pupil and teacher stress, create a culture of fear, tend to drive teachers out of the profession, lead to various gaming behaviours (e.g., excluding pupils from tests and cheating), and that the published scores are often presented with insufficient guidance, caveats, or quantification of statistical uncertainty. Perhaps, most worryingly, there is still very little research demonstrating the actual improvement to pupil learning that school accountability via pupil test scores and school value-added measures are meant to bring about (NFER, 2018).

Our own view is that the results presented here, coupled with these more general concerns, raise serious doubts about not just Progress 8, but test-based school accountability more generally. In terms of Progress 8, the types of automated data driven decision making that the Government currently aspires to, whereby schools falling below a single floor standard are declared underperforming, cannot be supported by the data. Our view is that, for school accountability purposes, the most school value-added measures can be used for is as 'screening devices' to choose schools for careful sensitive further investigation (Foley and Goldstein, 2012). However, we believe that a better use is simply as tools for school self-evaluation where they can potentially help inform schools on the policies and practices which help different pupil groups to reach their potential, but further discussion of this is outside the scope of the present article.

Table 1.

Pupil- and school-level summary statistics for, Progress 8 and Adjusted Progress 8.

| Description | Mean | SD | Min | 10th | 25th | 50th | 75th | 90th | Max |
|---|---|---|---|---|---|---|---|---|---|
| Pupils (N = 502,851) | | | | | | | | | |
| Progress 8 | 0.00 | 1.06 | -7.39 | -1.25 | -0.52 | 0.11 | 0.69 | 1.18 | 5.57 |
| Adjusted Progress 8 | 0.00 | 0.99 | -7.34 | -1.17 | -0.51 | 0.09 | 0.63 | 1.11 | 5.44 |
| Schools (N = 3,098) | | | | | | | | | |
| Progress 8 | -0.03 | 0.40 | -3.54 | -0.50 | -0.23 | 0.00 | 0.24 | 0.43 | 1.37 |
| Adjusted Progress 8 | -0.01 | 0.35 | -3.19 | -0.40 | -0.20 | 0.01 | 0.20 | 0.38 | 1.30 |

Note.

10th, 25th, 50th, 75th, and 90th denote percentiles of the relevant score distributions.

Table 2.

School Progress 8 and school Adjusted Progress 8 bandings.

| Banding | Definition | | Number and % of schools | |
|---|---|---|---|---|
| | Score | Significant | Progress 8 | Adjusted Progress 8 |
| 5 = Well above average | $\geq 0.5$ | Yes | 193 (6.2%) | 148 (4.8%) |
| 4 = Above average | $> 0$ & $< 0.5$ | Yes | 764 (24.7%) | 783 (25.3%) |
| 3 = Average | | No | 1213 (39.2%) | 1278 (41.3%) |
| 2 = Below average | $\geq -0.5$ & $< 0$ | Yes | 625 (20.2%) | 693 (22.4%) |
| 1 = Well below average | $< -0.5$ | Yes | 303 (9.8%) | 196 (6.3%) |

Note.

Definitions reproduced from DfE (2018b).

Significant = whether the score is significantly different from 0.

Number of schools = 3,098.

Table 3.

Cross-tabulation of school Progress 8 bandings by school Adjusted Progress 8 bandings.

|                    | Adjusted Progress 8 banding | | | | | |
|--------------------|------------|-------|---------|-------|------------|-------|
| Progress 8 banding | Well below | Below | Average | Above | Well above | Total |
| Well above | 0    | 0     | 5     | 101   | 87    | 193   |
|            | 0.0% | 0.0%  | 2.6%  | 52.3% | 45.1% | 100%  |
| Above      | 0    | 3     | 195   | 511   | 55    | 764   |
|            | 0.0% | 0.4%  | 25.5% | 66.9% | 7.2%  | 100%  |
| Average    | 0    | 141   | 898   | 168   | 6     | 1,213 |
|            | 0.0% | 11.6% | 74.0% | 13.9% | 0.5%  | 100%  |
| Below      | 16   | 434   | 172   | 3     | 0     | 625   |
|            | 2.6% | 69.4% | 27.5% | 0.5%  | 0.0%  | 100%  |
| Well below | 180  | 115   | 8     | 0     | 0     | 303   |
|            | 59.4%| 38.0% | 2.6%  | 0.0%  | 0.0%  | 100%  |
| Total      | 196  | 963   | 1,278 | 783   | 148   | 3,098 |
|            | 6.3% | 22.4% | 41.3% | 25.3% | 4.8%  | 100%  |

Note.

Definitions of bandings are giving in Table 3.

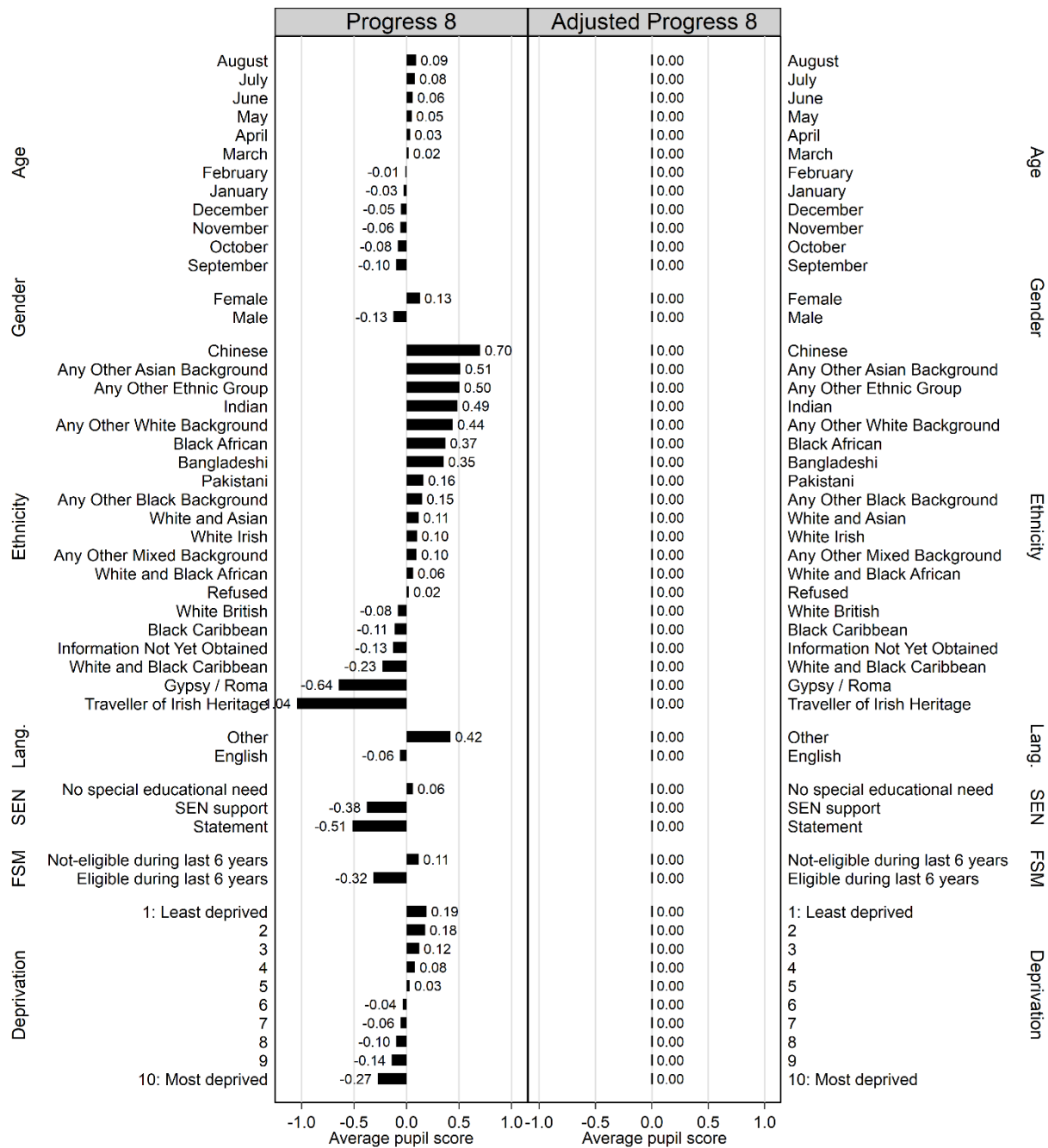

Figure 1.

Average pupil Progress 8 and Adjusted Progress 8 scores by pupil characteristics.

Note.

By definition, there is no variation in average Adjusted Progress 8 by pupil characteristic.

The number of pupils by pupil characteristic are given in Table S2.

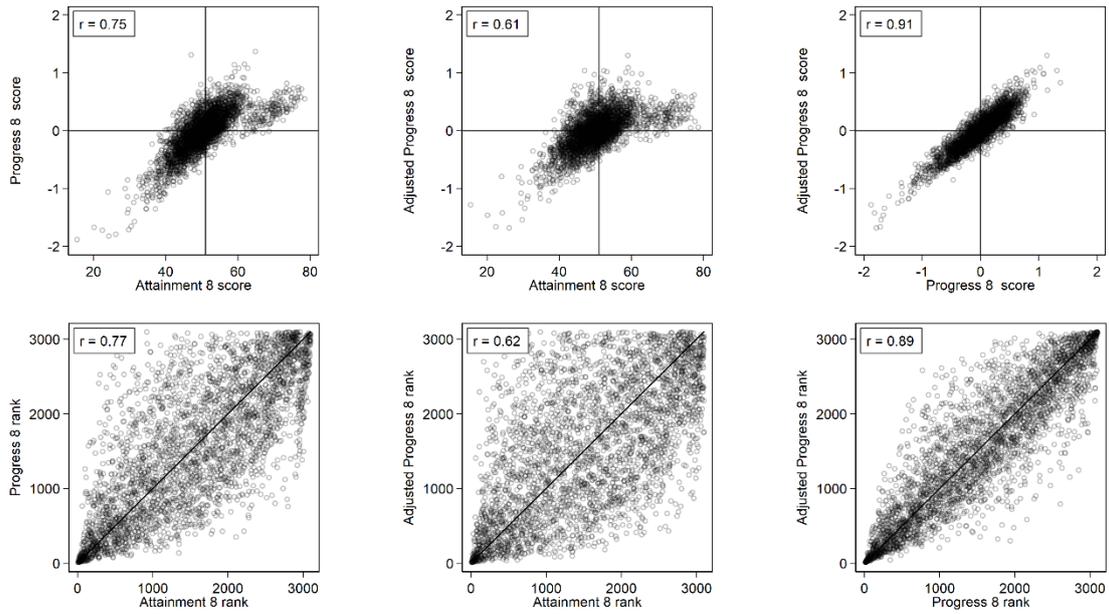

Figure 2.

Scatterplots of school average Attainment 8, Progress 8, and Adjusted Progress 8 scores (first row) and ranks (second row) with Pearson and Spearman rank correlations.

Note.

The horizontal and vertical lines in the first row of plots denote the mean values of the relevant variables.

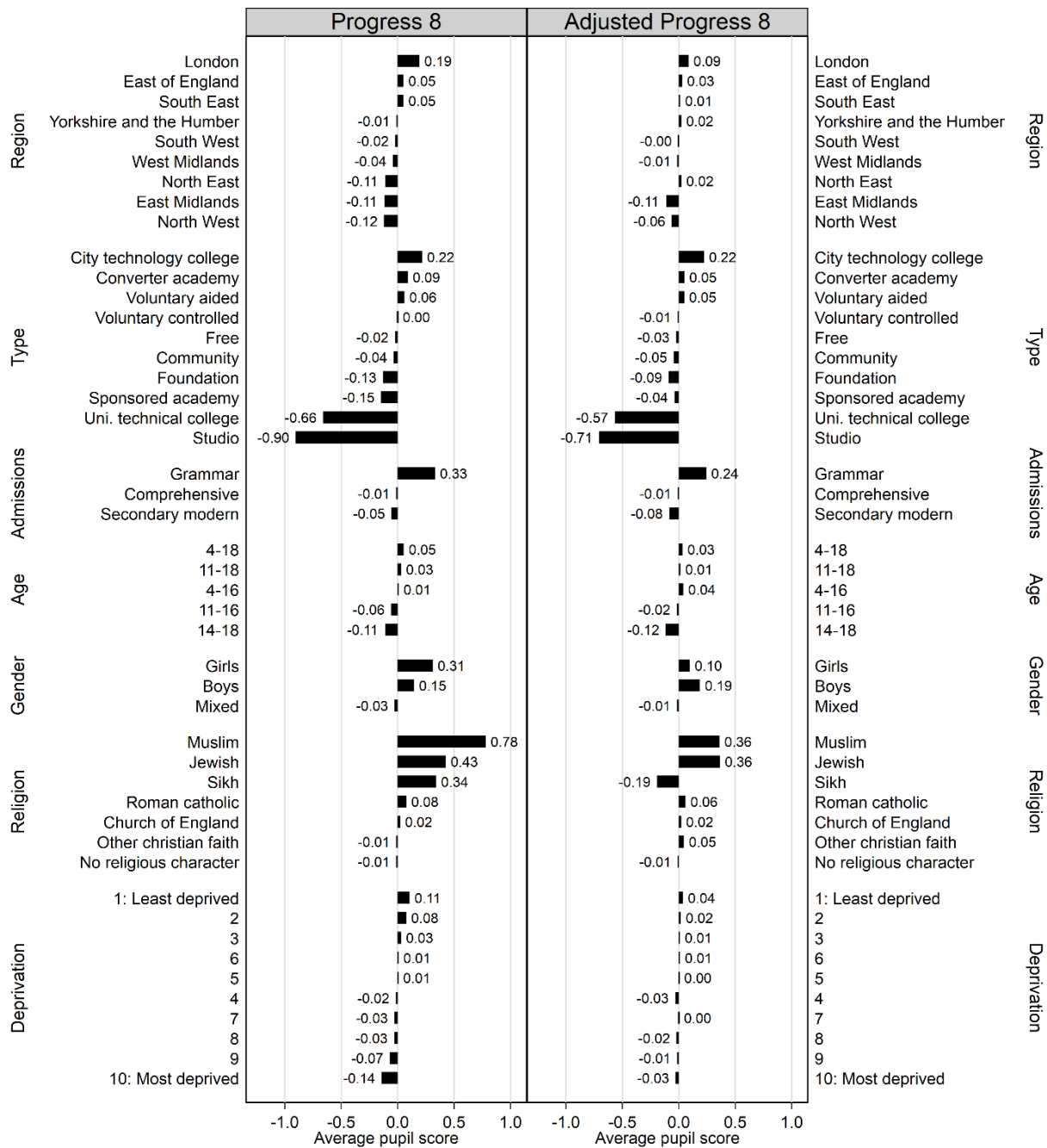

Figure 3.

Average pupil Progress 8 and Adjusted Progress 8 scores by school characteristics.

Note.

The categories of each school characteristic are sorted by average pupil Progress 8 score.

There are only three City technology colleges.

There is only one Sikh school and eight Muslim schools.

The number of pupils and schools by school characteristic are given in Table S3.

**Supporting Information**

Table S1.

Pupil- and school-level summa.ry statistics for Key Stage 2, Attainment 8, Progress 8 and Adjusted Progress 8

| Description | Mean | SD | Min | 10th | 25th | 50th | 75th | 90th | Max |
|---|---|---|---|---|---|---|---|---|---|
| \multicolumn{10}{c}{Pupils (N = 502,851)} | | | | | | | | | |
| Key Stage 2 | 4.60 | 0.70 | 1.50 | 3.70 | 4.20 | 4.70 | 5.10 | 5.40 | 5.80 |
| Attainment 8 | 51.02 | 16.16 | 0.00 | 29.00 | 42.00 | 53.00 | 62.00 | 70.00 | 86.00 |
| Progress 8 | 0.00 | 1.06 | -7.39 | -1.25 | -0.52 | 0.11 | 0.69 | 1.18 | 5.57 |
| Adjusted Progress 8 | 0.00 | 0.99 | -7.34 | -1.17 | -0.51 | 0.09 | 0.63 | 1.11 | 5.44 |
| \multicolumn{10}{c}{Schools (N = 3,098)} | | | | | | | | | |
| Key Stage 2 | 4.58 | 0.27 | 3.40 | 4.28 | 4.42 | 4.56 | 4.71 | 4.84 | 5.66 |
| Attainment 8 | 50.48 | 7.47 | 5.36 | 42.45 | 46.12 | 50.24 | 54.05 | 58.55 | 78.43 |
| Progress 8 | -0.03 | 0.40 | -3.54 | -0.50 | -0.23 | 0.00 | 0.24 | 0.43 | 1.37 |
| Adjusted Progress 8 | -0.01 | 0.35 | -3.19 | -0.40 | -0.20 | 0.01 | 0.20 | 0.38 | 1.30 |

Note.

10th, 25th, 50th, 75th, and 90th denote percentiles of the relevant score distributions.

Table S2.

Distribution of pupils by pupil demographic and socioeconomic characteristics.

| Variable | Pupils | | KS2 | A8 | P8 | AP8 |
| --- | --- | --- | --- | --- | --- | --- |
| | N | % | | | | |
| KS2 prior attainment group | | | | | | |
| 1: Lowest | 960 | 0.2 | 1.50 | 14.5 | 0.0 | 0.0 |
| 2 | 1164 | 0.2 | 2.00 | 20.1 | 0.0 | 0.0 |
| 3 | 7692 | 1.5 | 2.50 | 21.3 | 0.0 | 0.0 |
| 4 | 3133 | 0.6 | 2.80 | 22.5 | 0.0 | 0.0 |
| 5 | 2413 | 0.5 | 2.90 | 24.6 | 0.0 | 0.0 |
| 6 | 2417 | 0.5 | 3.00 | 25.4 | 0.0 | 0.0 |
| 7 | 3287 | 0.7 | 3.10 | 26.5 | 0.0 | 0.0 |
| 8 | 3359 | 0.7 | 3.20 | 27.6 | 0.0 | 0.0 |
| 9 | 4757 | 0.9 | 3.30 | 29.3 | 0.0 | 0.0 |
| 10 | 5228 | 1.0 | 3.40 | 30.1 | 0.0 | 0.0 |
| 11 | 6357 | 1.3 | 3.50 | 31.5 | 0.0 | 0.0 |
| 12 | 7499 | 1.5 | 3.60 | 33.1 | 0.0 | 0.0 |
| 13 | 8337 | 1.7 | 3.70 | 34.6 | 0.0 | 0.0 |
| 14 | 10041 | 2.0 | 3.80 | 36.1 | 0.0 | 0.0 |
| 15 | 12033 | 2.4 | 3.90 | 38.0 | 0.0 | 0.0 |
| 16 | 13679 | 2.7 | 4.00 | 39.4 | 0.0 | 0.0 |
| 17 | 16026 | 3.2 | 4.10 | 41.0 | 0.0 | 0.0 |
| 18 | 19589 | 3.9 | 4.20 | 42.7 | 0.0 | 0.0 |
| 19 | 23473 | 4.7 | 4.30 | 44.5 | 0.0 | 0.0 |
| 20 | 25852 | 5.1 | 4.40 | 46.2 | 0.0 | 0.0 |
| 21 | 29549 | 5.9 | 4.50 | 47.9 | 0.0 | 0.0 |
| 22 | 30450 | 6.1 | 4.60 | 50.0 | 0.0 | 0.0 |
| 23 | 30669 | 6.1 | 4.70 | 51.9 | 0.0 | 0.0 |
| 24 | 31371 | 6.2 | 4.80 | 53.9 | 0.0 | 0.0 |
| 25 | 30990 | 6.2 | 4.90 | 55.8 | 0.0 | 0.0 |
| 26 | 29952 | 6.0 | 5.00 | 57.7 | 0.0 | 0.0 |
| 27 | 28983 | 5.8 | 5.10 | 59.9 | 0.0 | 0.0 |
| 28 | 27346 | 5.4 | 5.20 | 62.0 | 0.0 | 0.0 |
| 29 | 24938 | 5.0 | 5.30 | 64.1 | 0.0 | 0.0 |
| 30 | 21913 | 4.4 | 5.40 | 66.5 | 0.0 | 0.0 |
| 31 | 18167 | 3.6 | 5.50 | 68.8 | 0.0 | 0.0 |
| 32 | 12225 | 2.4 | 5.60 | 71.5 | 0.0 | 0.0 |
| 33 | 6505 | 1.3 | 5.70 | 73.9 | 0.0 | 0.0 |
| 34: Highest | 2497 | 0.5 | 5.80 | 76.1 | 0.0 | 0.0 |
| Month of birth | | | | | | |
| September | 43346 | 8.6 | 4.73 | 52.4 | -0.10 | 0.00 |
| October | 41981 | 8.3 | 4.72 | 52.3 | -0.08 | 0.00 |
| November | 41113 | 8.2 | 4.68 | 51.9 | -0.06 | 0.00 |
| December | 42700 | 8.5 | 4.65 | 51.4 | -0.05 | 0.00 |
| January | 42124 | 8.4 | 4.62 | 51.1 | -0.03 | 0.00 |
| February | 38949 | 7.7 | 4.60 | 51.0 | -0.01 | 0.00 |
| March | 42158 | 8.4 | 4.60 | 51.1 | 0.02 | 0.00 |
| April | 40458 | 8.0 | 4.57 | 50.7 | 0.03 | 0.00 |
| May | 42601 | 8.5 | 4.54 | 50.5 | 0.05 | 0.00 |
| June | 40983 | 8.2 | 4.52 | 50.1 | 0.06 | 0.00 |
| July | 43493 | 8.6 | 4.50 | 49.9 | 0.08 | 0.00 |
| August | 42945 | 8.5 | 4.47 | 49.6 | 0.09 | 0.00 |
| Gender | | | | | | |
| Male | 253733 | 50.5 | 4.56 | 49.1 | -0.13 | 0.00 |
| Female | 249118 | 49.5 | 4.64 | 53.0 | 0.13 | 0.00 |
| Ethnicity | | | | | | |
| White British | 380949 | 75.8 | 4.62 | 50.5 | -0.08 | 0.00 |
| White Irish | 1606 | 0.3 | 4.79 | 55.5 | 0.10 | 0.00 |
| Traveller of Irish Heritage | 104 | 0.0 | 4.04 | 31.5 | -1.04 | 0.00 |
| Gypsy / Roma | 659 | 0.1 | 3.38 | 26.6 | -0.65 | 0.00 |
| Any Other White Background | 17129 | 3.4 | 4.46 | 53.2 | 0.44 | 0.00 |
| Black African | 14379 | 2.9 | 4.47 | 52.5 | 0.37 | 0.00 |
| Black Caribbean | 6650 | 1.3 | 4.42 | 46.4 | -0.11 | 0.00 |
| Any Other Black Background | 2690 | 0.5 | 4.42 | 49.2 | 0.15 | 0.00 |
| Indian | 12426 | 2.5 | 4.75 | 58.6 | 0.49 | 0.00 |
| Pakistani | 18722 | 3.7 | 4.42 | 49.5 | 0.16 | 0.00 |
| Bangladeshi | 7709 | 1.5 | 4.52 | 53.2 | 0.35 | 0.00 |
| Any Other Asian Background | 6900 | 1.4 | 4.67 | 57.6 | 0.51 | 0.00 |
| Chinese | 1585 | 0.3 | 4.95 | 64.8 | 0.70 | 0.00 |
| White and Black African | 2390 | 0.5 | 4.58 | 51.3 | 0.06 | 0.00 |
| White and Black Caribbean | 6873 | 1.4 | 4.53 | 47.3 | -0.23 | 0.00 |
| White and Asian | 4656 | 0.9 | 4.75 | 55.1 | 0.11 | 0.00 |
| Any Other Mixed Background | 6983 | 1.4 | 4.65 | 53.0 | 0.10 | 0.00 |
| Any Other Ethnic Group | 6198 | 1.2 | 4.45 | 53.8 | 0.50 | 0.00 |
| Information Not Yet Obtained | 2098 | 0.4 | 4.54 | 48.7 | -0.13 | 0.00 |
| Refused | 2145 | 0.4 | 4.63 | 51.8 | 0.02 | 0.00 |
| English as additional language | | | | | | |
| English as first language | 438585 | 87.2 | 4.62 | 50.8 | -0.06 | 0.00 |
| English as additional language | 64266 | 12.8 | 4.45 | 52.8 | 0.42 | 0.00 |
| Special educational needs | | | | | | |
| No special educational need | 436229 | 86.8 | 4.71 | 53.5 | 0.06 | 0.00 |
| SEN support | 55601 | 11.1 | 3.95 | 36.3 | -0.38 | 0.00 |
| Statement | 11021 | 2.2 | 3.48 | 29.0 | -0.52 | 0.00 |
| Free school meal status | | | | | | |
| Not-eligible during last 6 years | 369147 | 73.4 | 4.70 | 54.0 | 0.11 | 0.00 |
| Eligible during last 6 years | 133704 | 26.6 | 4.32 | 42.9 | -0.32 | 0.00 |
| Deprivation | | | | | | |
| 1: Least deprived | 50289 | 10.0 | 4.83 | 57.2 | 0.19 | 0.00 |
| 2 | 51790 | 10.3 | 4.77 | 55.9 | 0.18 | 0.00 |
| 3 | 49086 | 9.8 | 4.73 | 54.5 | 0.12 | 0.00 |
| 4 | 51072 | 10.2 | 4.68 | 53.3 | 0.08 | 0.00 |
| 5 | 50340 | 10.0 | 4.63 | 51.8 | 0.03 | 0.00 |
| 6 | 49321 | 9.8 | 4.58 | 50.2 | -0.04 | 0.00 |
| 7 | 50172 | 10.0 | 4.51 | 48.9 | -0.06 | 0.00 |
| 8 | 50853 | 10.1 | 4.46 | 47.6 | -0.10 | 0.00 |
| 9 | 49761 | 9.9 | 4.42 | 46.5 | -0.14 | 0.00 |
| 10: Most deprived | 50167 | 10.0 | 4.37 | 44.3 | -0.27 | 0.00 |

Note.

Sample size = 502,851 pupils in 3,098 schools.

A8 = Attainment 8.

KS2 = Key stage 2.

P8 = Progress 8.

AP8 = Adjusted Progress 8.

Table S3.

Distribution of pupils and schools by school characteristics.

| Characteristic | Pupils | | Schools | | KS2 | A8 | P8 | AP8 |
|---|---|---|---|---|---|---|---|---|
| | N | % | N | % | | | | |
| Region | | | | | | | | |
| London | 68696 | 13.7 | 431 | 13.9 | 4.6 | 53.3 | 0.19 | 0.09 |
| South East | 79504 | 15.8 | 474 | 15.3 | 4.6 | 51.9 | 0.05 | 0.01 |
| South West | 49350 | 9.8 | 309 | 10.0 | 4.6 | 51.0 | -0.02 | -0.00 |
| West Midlands | 56529 | 11.2 | 373 | 12.0 | 4.6 | 50.2 | -0.04 | -0.01 |
| North West | 69954 | 13.9 | 447 | 14.4 | 4.6 | 50.3 | -0.12 | -0.06 |
| North East | 24827 | 4.9 | 152 | 4.9 | 4.6 | 50.0 | -0.11 | 0.02 |
| Yorkshire & Humber | 51908 | 10.3 | 298 | 9.6 | 4.5 | 49.8 | -0.01 | 0.02 |
| East Midlands | 44618 | 8.9 | 269 | 8.7 | 4.6 | 49.8 | -0.11 | -0.11 |
| East of England | 57465 | 11.4 | 345 | 11.1 | 4.6 | 51.2 | 0.05 | 0.03 |
| School type | | | | | | | | |
| Community | 89762 | 17.9 | 538 | 17.4 | 4.6 | 50.1 | -0.04 | -0.05 |
| Foundation | 44843 | 8.9 | 275 | 8.9 | 4.5 | 48.2 | -0.13 | -0.09 |
| Voluntary aided | 40932 | 8.1 | 273 | 8.8 | 4.7 | 52.6 | 0.06 | 0.05 |
| Voluntary controlled | 6274 | 1.2 | 34 | 1.1 | 4.6 | 51.2 | 0.00 | -0.01 |
| City tech. college | 516 | 0.1 | 3 | 0.1 | 4.9 | 58.7 | 0.22 | 0.22 |
| Sponsored academy | 79352 | 15.8 | 560 | 18.1 | 4.4 | 46.3 | -0.15 | -0.04 |
| Converter academy | 236215 | 47.0 | 1320 | 42.6 | 4.7 | 53.4 | 0.09 | 0.05 |
| Free | 1649 | 0.3 | 27 | 0.9 | 4.6 | 50.7 | -0.02 | -0.03 |
| Studio | 1046 | 0.2 | 30 | 1.0 | 4.3 | 36.4 | -0.90 | -0.71 |
| Uni. tech. college | 1765 | 0.4 | 26 | 0.8 | 4.5 | 43.0 | -0.66 | -0.57 |

| | | | | | | | |
|---|---|---|---|---|---|---|---|
| Further ed. college | 497 | 0.1 | 12 | 0.4 | 4.2 | 25.2 | -1.82 | -1.39 |
| School admissions | | | | | | | | |
| Comprehensive | 464874 | 92.4 | 2819 | 91.0 | 4.6 | 50.3 | -0.01 | -0.01 |
| Grammar | 20472 | 4.1 | 162 | 5.2 | 5.4 | 69.0 | 0.33 | 0.24 |
| Secondary modern | 17505 | 3.5 | 117 | 3.8 | 4.5 | 47.9 | -0.05 | -0.08 |
| Age range | | | | | | | | |
| 11-18 | 321951 | 64.0 | 1881 | 60.7 | 4.6 | 52.0 | 0.03 | 0.01 |
| 11-16 | 146864 | 29.2 | 971 | 31.3 | 4.5 | 49.3 | -0.06 | -0.02 |
| 14-18 | 18694 | 3.7 | 135 | 4.4 | 4.5 | 48.9 | -0.11 | -0.12 |
| 4-18 | 12086 | 2.4 | 83 | 2.7 | 4.5 | 49.8 | 0.05 | 0.03 |
| 4-16 | 3256 | 0.6 | 28 | 0.9 | 4.4 | 47.8 | 0.01 | 0.04 |
| School gender | | | | | | | | |
| Mixed | 452401 | 90.0 | 2738 | 88.4 | 4.6 | 50.2 | -0.03 | -0.01 |
| Boys | 20096 | 4.0 | 151 | 4.9 | 4.9 | 58.0 | 0.15 | 0.19 |
| Girls | 30354 | 6.0 | 209 | 6.7 | 4.8 | 58.5 | 0.31 | 0.10 |
| School religion | | | | | | | | |
| None | 414268 | 82.4 | 2524 | 81.5 | 4.6 | 50.7 | -0.01 | -0.01 |
| Church of England | 27687 | 5.5 | 176 | 5.7 | 4.6 | 51.7 | 0.02 | 0.02 |
| Roman catholic | 48710 | 9.7 | 310 | 10.0 | 4.7 | 52.9 | 0.08 | 0.06 |
| Other Christian faith | 10217 | 2.0 | 68 | 2.2 | 4.6 | 50.8 | -0.01 | 0.05 |
| Jewish | 1161 | 0.2 | 11 | 0.4 | 4.9 | 60.5 | 0.43 | 0.36 |
| Muslim | 638 | 0.1 | 8 | 0.3 | 4.6 | 59.4 | 0.78 | 0.36 |
| Sikh | 170 | 0.0 | 1 | 0.0 | 4.8 | 57.8 | 0.34 | -0.19 |
| School IDACI decile | | | | | | | | |
| 1: Least deprived | 52329 | 10.4 | 288 | 9.3 | 4.7 | 54.2 | 0.11 | 0.04 |
| 2 | 58476 | 11.6 | 329 | 10.6 | 4.7 | 53.3 | 0.08 | 0.02 |
| 3 | 52965 | 10.5 | 313 | 10.1 | 4.7 | 52.3 | 0.03 | 0.01 |
| 4 | 50085 | 10.0 | 303 | 9.8 | 4.6 | 51.0 | -0.02 | -0.03 |
| 5 | 54182 | 10.8 | 325 | 10.5 | 4.6 | 51.3 | 0.01 | 0.00 |
| 6 | 52979 | 10.5 | 332 | 10.7 | 4.6 | 51.0 | 0.01 | 0.01 |
| 7 | 52004 | 10.3 | 327 | 10.6 | 4.6 | 50.0 | -0.03 | 0.00 |
| 8 | 50528 | 10.0 | 320 | 10.3 | 4.5 | 49.2 | -0.03 | -0.02 |
| 9 | 41498 | 8.3 | 289 | 9.3 | 4.5 | 48.3 | -0.07 | -0.01 |
| 10: Most deprived | 37805 | 7.5 | 272 | 8.8 | 4.5 | 47.5 | -0.14 | -0.03 |

Note.

Sample size = 502,851 pupils in 3,098 schools.

KS2 = Key stage 2.

A8 = Attainment 8.

P8 = Progress 8.

AP8 = Adjusted Progress 8.

Table S4.

Model results for Progress 8 and Adjusted Progress 8 linear regression models.

| Variable | Progress 8 | | Adjusted Progress 8 | |
|---|---|---|---|---|
| | Coef. | SE | Coef. | SE |
| Constant | 14.52 | 0.55 | 19.74 | 0.33 |
| KS2 group (ref. cat. = KS2 Group 1) | | | | |
| KS2 group 2 | 5.55 | 0.66 | 5.52 | 0.43 |
| KS2 group 3 | 6.73 | 0.54 | 6.73 | 0.34 |
| KS2 group 4 | 8.00 | 0.57 | 7.71 | 0.37 |
| KS2 group 5 | 10.11 | 0.60 | 9.29 | 0.38 |
| KS2 group 6 | 10.83 | 0.60 | 9.86 | 0.38 |
| KS2 group 7 | 11.94 | 0.57 | 10.84 | 0.37 |
| KS2 group 8 | 13.11 | 0.58 | 11.67 | 0.37 |
| KS2 group 9 | 14.78 | 0.57 | 13.04 | 0.35 |
| KS2 group 10 | 15.62 | 0.57 | 13.63 | 0.35 |
| KS2 group 11 | 16.97 | 0.56 | 14.75 | 0.35 |
| KS2 group 12 | 18.62 | 0.56 | 16.03 | 0.34 |
| KS2 group 13 | 20.04 | 0.56 | 17.22 | 0.34 |
| KS2 group 14 | 21.56 | 0.56 | 18.48 | 0.34 |
| KS2 group 15 | 23.47 | 0.56 | 20.09 | 0.34 |
| KS2 group 16 | 24.83 | 0.56 | 21.24 | 0.33 |
| KS2 group 17 | 26.43 | 0.55 | 22.72 | 0.33 |
| KS2 group 18 | 28.16 | 0.55 | 24.18 | 0.33 |
| KS2 group 19 | 29.94 | 0.55 | 25.86 | 0.33 |
| KS2 group 20 | 31.70 | 0.55 | 27.38 | 0.33 |
| KS2 group 21 | 33.34 | 0.55 | 28.89 | 0.33 |
| KS2 group 22 | 35.43 | 0.55 | 30.76 | 0.33 |
| KS2 group 23 | 37.33 | 0.55 | 32.53 | 0.33 |
| KS2 group 24 | 39.39 | 0.55 | 34.40 | 0.33 |
| KS2 group 25 | 41.32 | 0.55 | 36.18 | 0.33 |
| KS2 group 26 | 43.17 | 0.55 | 37.87 | 0.33 |
| KS2 group 27 | 45.40 | 0.55 | 39.94 | 0.33 |
| KS2 group 28 | 47.51 | 0.55 | 41.92 | 0.33 |
| KS2 group 29 | 49.62 | 0.55 | 43.93 | 0.33 |
| KS2 group 30 | 52.01 | 0.55 | 46.11 | 0.33 |
| KS2 group 31 | 54.30 | 0.56 | 48.27 | 0.33 |
| KS2 group 32 | 56.96 | 0.56 | 50.69 | 0.34 |
| KS2 group 33 | 59.34 | 0.56 | 52.90 | 0.35 |
| KS2 group 34 | 61.54 | 0.56 | 54.91 | 0.38 |
| Month of birth (ref. cat. = September) | | | | |
| October | | | 0.15 | 0.07 |
| November | | | 0.35 | 0.07 |
| December | | | 0.42 | 0.07 |
| January | | | 0.59 | 0.07 |
| February | | | 0.78 | 0.07 |
| March | | | 0.99 | 0.07 |
| April | | | 1.12 | 0.07 |
| May | | | 1.21 | 0.07 |
| June | | | 1.30 | 0.07 |

| | | |
|---|---:|---:|
| July | 1.49 | 0.07 |
| August | 1.62 | 0.07 |
| **Gender (ref. cat. = Male)** | | |
| Female | 2.44 | 0.03 |
| **Ethnicity (ref. cat. = White British)** | | |
| White Irish | 2.02 | 0.25 |
| Traveller of Irish Heritage | -6.92 | 0.97 |
| Gypsy / Roma | -5.63 | 0.39 |
| Any Other White Background | 3.90 | 0.09 |
| Black African | 5.42 | 0.09 |
| Black Caribbean | 1.80 | 0.12 |
| Any Other Black Background | 3.75 | 0.19 |
| Indian | 4.16 | 0.10 |
| Pakistani | 1.93 | 0.09 |
| Bangladeshi | 4.49 | 0.13 |
| Any Other Asian Background | 4.71 | 0.13 |
| Chinese | 6.26 | 0.25 |
| White and Black African | 2.46 | 0.20 |
| White and Black Caribbean | 0.04 | 0.12 |
| White and Asian | 2.08 | 0.15 |
| Any Other Mixed Background | 2.32 | 0.12 |
| Any Other Ethnic Group | 5.67 | 0.14 |
| Information Not Yet Obtained | -0.14 | 0.22 |
| Refused | 1.36 | 0.21 |
| **First language (ref. cat. English)** | | |

| | | | |
|---|---|---|---|
| Other | | 2.55 | 0.07 |
| SEN (ref. cat. = None) | | | |
| SEN support | | -4.42 | 0.05 |
| Statement | | -6.88 | 0.10 |
| Eligible for FSM (ref. cat. = No) | | | |
| Yes | | -4.01 | 0.04 |
| Deprivation (ref. cat. = decile 1) | | | |
| IDACI decile 2 | | -0.22 | 0.06 |
| IDACI decile 3 | | -0.79 | 0.06 |
| IDACI decile 4 | | -1.28 | 0.06 |
| IDACI decile 5 | | -1.87 | 0.06 |
| IDACI decile 6 | | -2.66 | 0.06 |
| IDACI decile 7 | | -2.99 | 0.06 |
| IDACI decile 8 | | -3.43 | 0.06 |
| IDACI decile 9 | | -3.82 | 0.07 |
| IDACI decile 10 | | -4.52 | 0.07 |
| Adjusted R-squared | 0.570 | 0.624 | |
| RMSE | 1.060 | 0.990 | |
| Number of schools | 3098 | 3098 | |
| Number of pupils | 502851 | 502851 | |

Note.

Coef. = Regression coefficient.

RMSE = Root mean squared error (the standard deviation of the pupil progress scores).

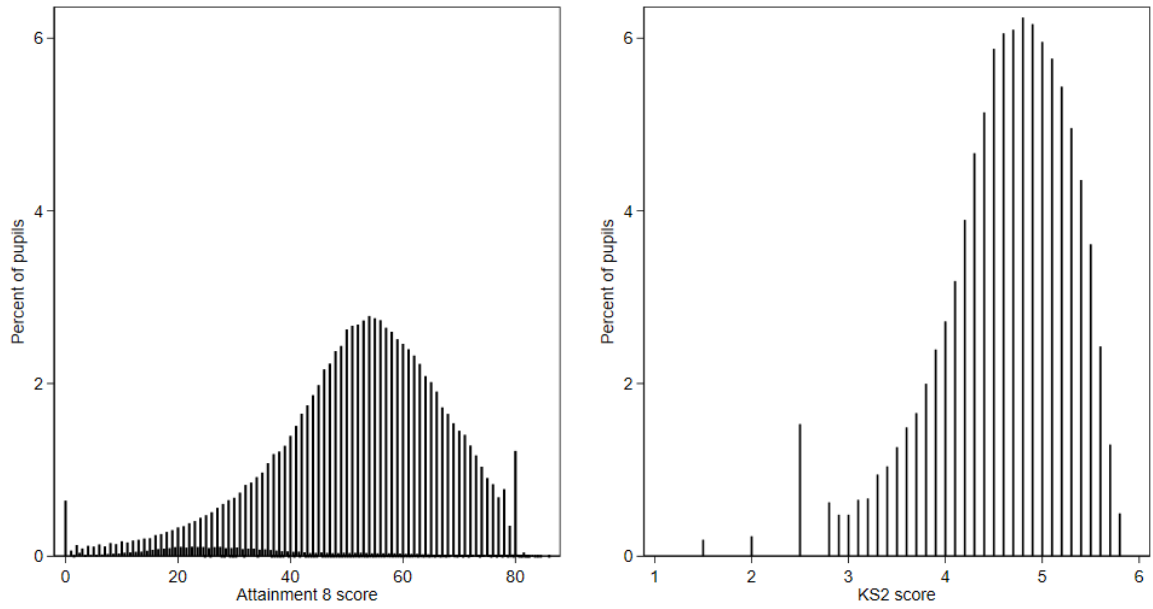

Figure S1.

Distribution of pupil Attainment 8 scores and pupil KS2 scores.

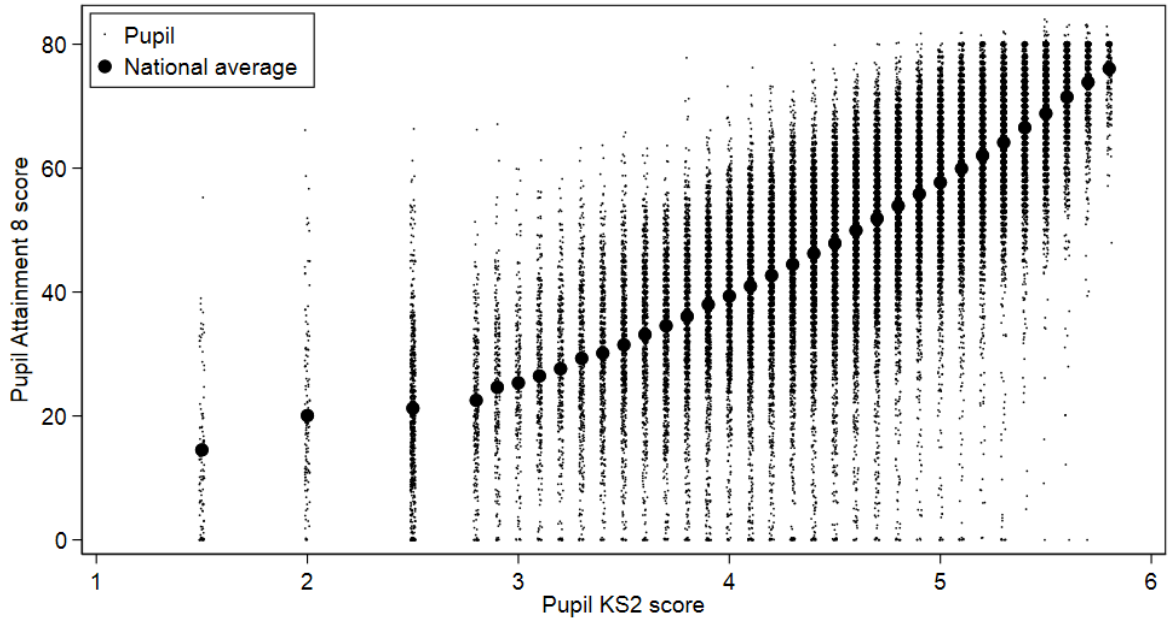

Figure S2.

Scatterplot of pupil Attainment 8 scores against pupil KS2 scores.

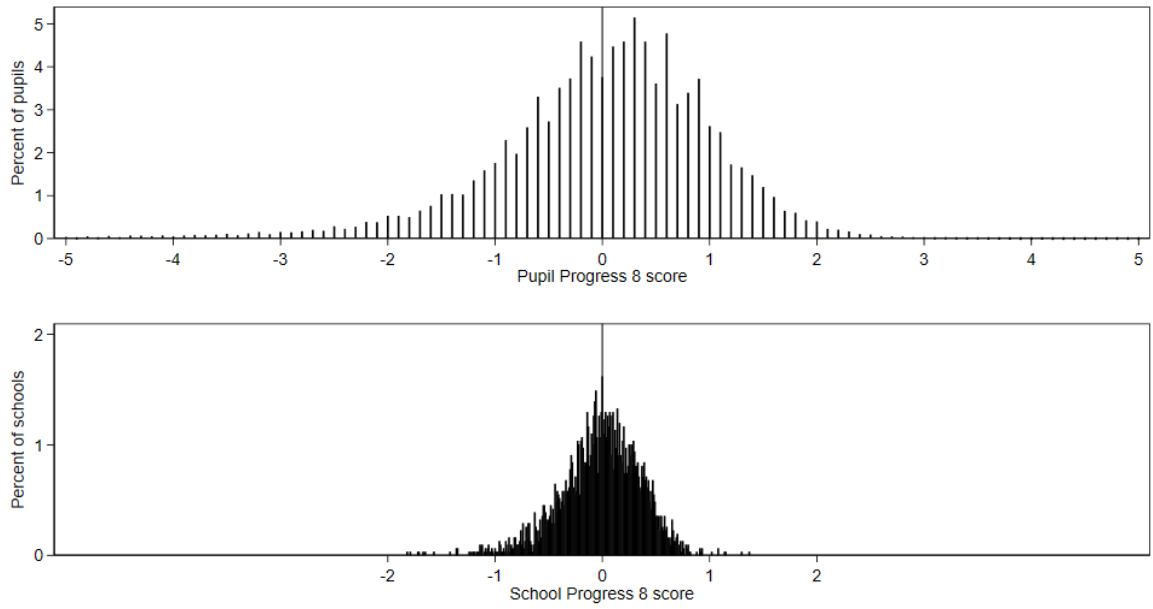

Figure S3.

Distribution of pupil Progress 8 and school Progress 8 scores.

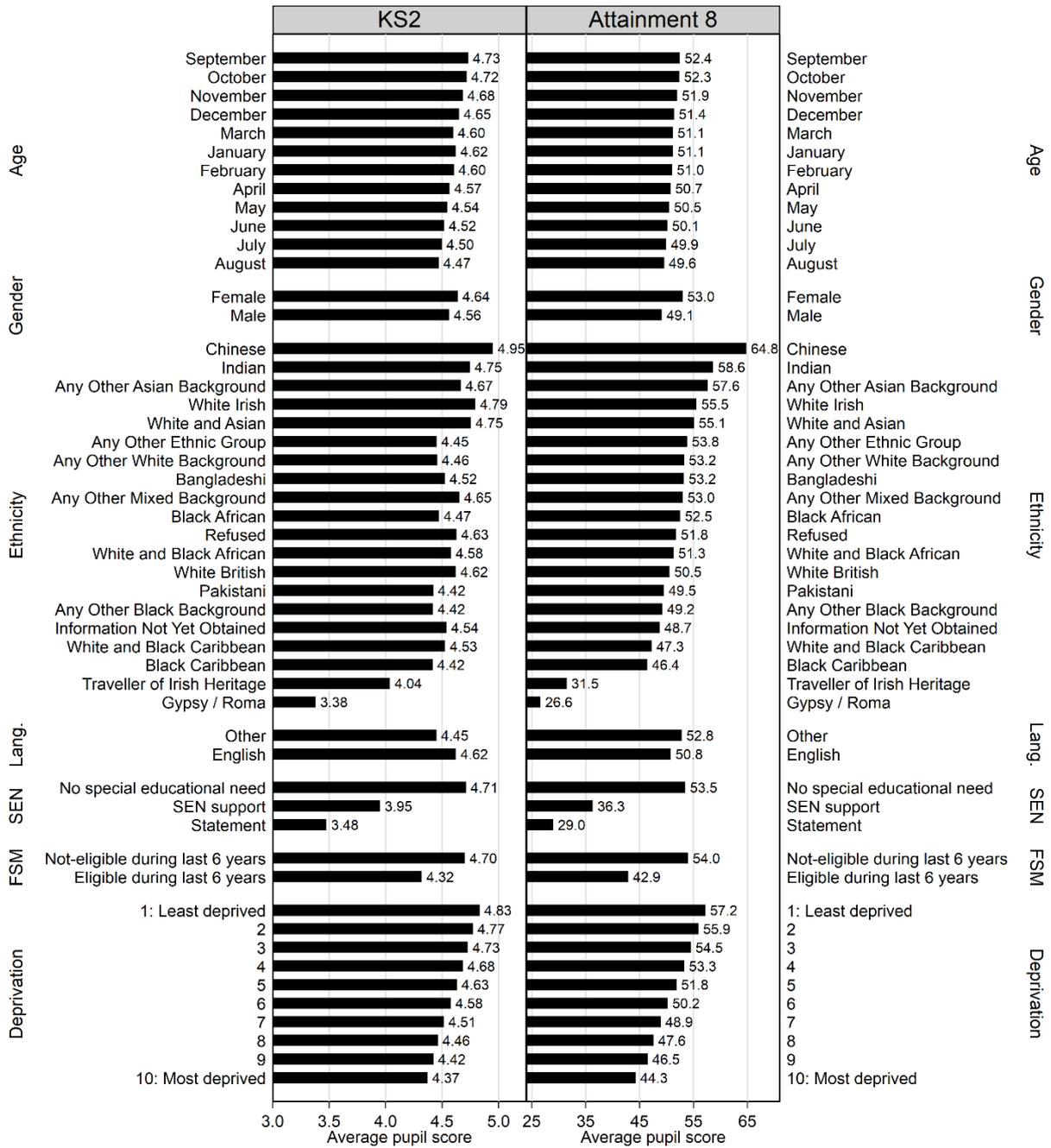

Figure S4.

Average pupil KS2 and Attainment 8 scores by pupil characteristics.

Note.

The categories of each pupil characteristic are sorted by average pupil Attainment 8 score.

The number of pupils by pupil characteristic are given in Table S2.

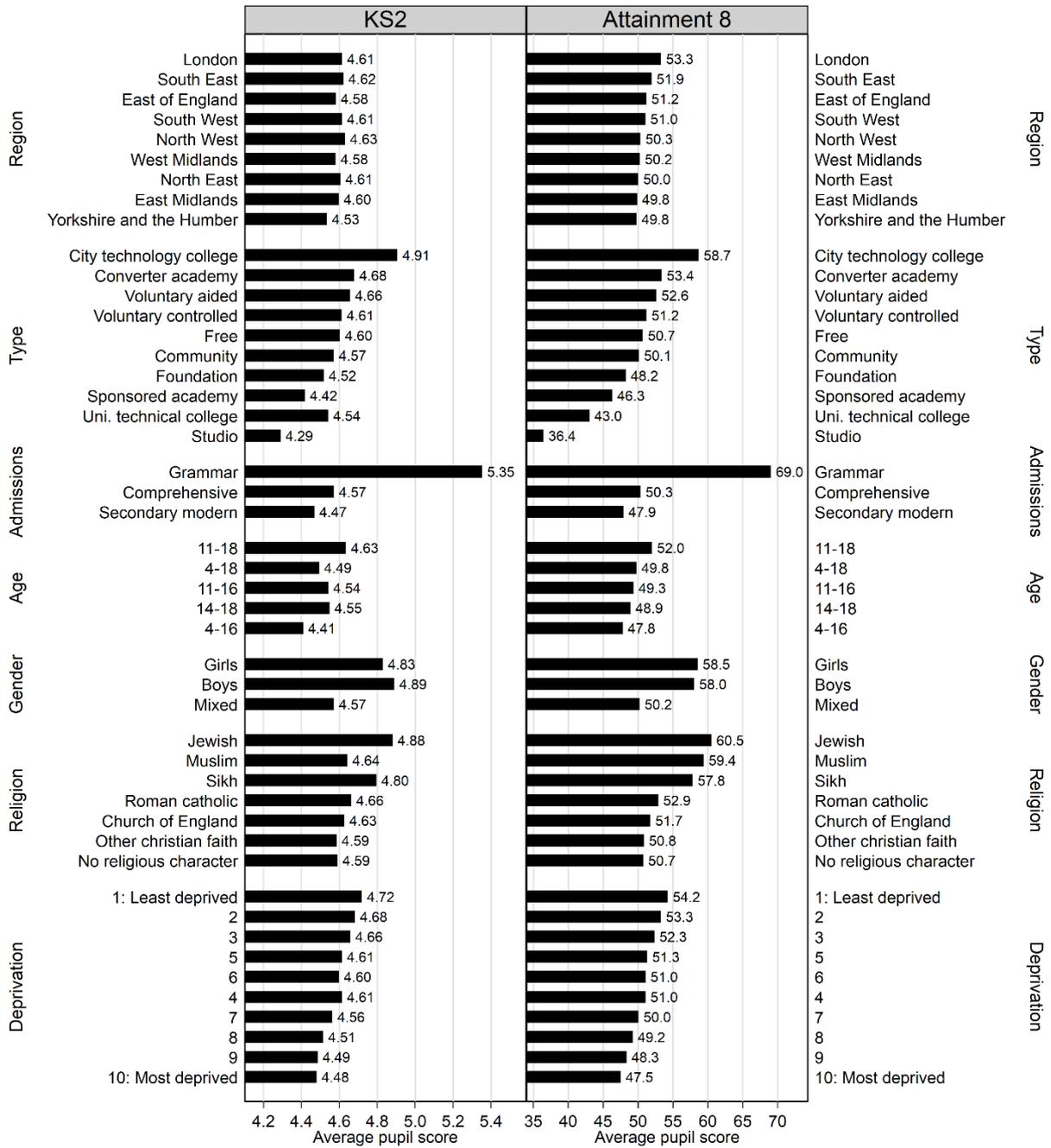

Figure S5.

Average pupil KS2 and Attainment 8 scores by school characteristics.

Note.

The categories of each school characteristic are sorted by average pupil Attainment 8 score.

The number of pupils and schools by school characteristic are given in Table S3.